\documentclass[twoside,11pt]{article} 

\usepackage{graphicx}
\usepackage{amsmath}
\usepackage{amssymb}
\usepackage{amsthm}
\usepackage{epsfig}
\usepackage{bbm}
\usepackage[numbers]{natbib}
\usepackage{fullpage, amsfonts, algorithm, algorithmic}
\usepackage{graphicx,color}
\usepackage{lmodern}

\usepackage{float}
\usepackage{multirow}
\usepackage{multicol}
\usepackage{array}
\usepackage[dvipsnames]{xcolor}
\usepackage{url}
\usepackage{caption}
\usepackage{subcaption}
\usepackage{hyperref}
\usepackage{rotating}

\begin{document}



\title{Estimating Censored Spatial-Temporal Demand with Applications to Shared Micromobility}
\author{Alice Paul  \\
alice\_paul@brown.edu, Brown University \\
 \\
  Kyran Flynn \\
  kyran\textunderscore flynn@brown.edu, Brown University \\
  \\
  Cassandra Overney \\
  overney@mit.edu, Massachusetts Institute of Technology
  }

\maketitle

\begin{abstract}
In shared micromobility networks, such as bike-share and scooter-share networks, using trip data to accurately estimate demand in docked and dockless systems is critical to analyzing how the system is operating, such as identifying the number of dissatisfied users, operational costs, and equity in access, especially for city officials. However, the distribution of available bikes affects the distribution of observed trips. Users may walk from an unobserved cell location to an available bike masking the true location of user demand, and users may look for a bike and not find one, which is unobserved user demand. 
In collaboration with city planners from Providence, R.I.,~we present a flexible and interpretable framework to estimate spatial-temporal demand as a spatial non-homogeneous Poisson process that explicitly models how users choose a bike, bridging the gap between the docked and dockless methodology. Further, we present computational experiments highlighting that our method provides more accurate estimates of demand when there is incomplete availability compared to previous methods, and we comment on the results of our algorithm on data from Providence's dockless scooter-share network. 
Our estimation algorithm is publicly available through an efficient and user-friendly application designed for other city planners and organizations to help inform system planning.
\end{abstract}

\section{Introduction} 
\label{sec:intro} 

Shared micromobility networks, including bike-share and scooter-share networks, are becoming an established part of urban environments. Access to these networks expands the modes of transportation available and enables connections to existing public transit. When checking in or checking out a bike or scooter, users directly impact the state of the system and the distribution of available vehicles. This leads to a wealth of individual-level data that city officials can use to extract usage patterns, understand how the population interacts with the network, and inform contracts with the available operators to serve the city better. 

\begin{figure}[h]
\begin{center}
\caption{Service regions for Providence's Shared Micromobility Program \citep{provMicro}.} \label{fig:service}
\includegraphics[scale=0.3]{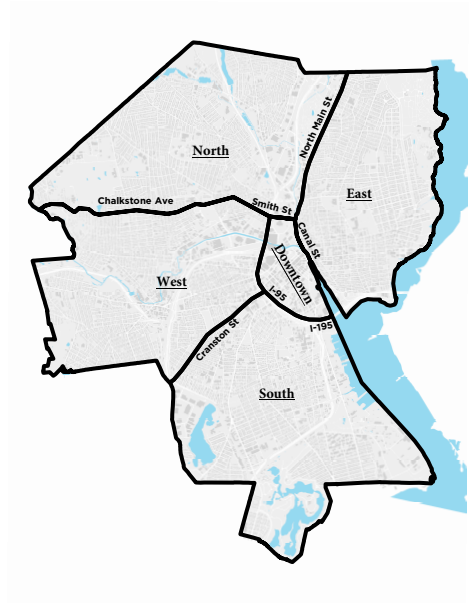}
\end{center}
\end{figure}

While initially shared micromobility networks had a docked structure in which users checked in and out bikes at a set of available stations, more recently, there has been an increase in dockless and hybrid systems in which users can return a bike or scooter anywhere by simply checking it back in through a mobile app and enabling the locking mechanism. Dockless systems increase the flexibility and efficiency of the network \citep{chen2020dockless,mooney2019freedom}, allowing the system's state to adapt to demand. In theory, this leads to increased equity in access. However, there are additional challenges as operators prioritize repositioning scooters or bikes to areas with low idle times over maintaining access across the service area \citep{chen2020dockless,mooney2019freedom}. This rebalancing makes it even more important for city officials to understand how availability affects observed demand to inform the regulations on operators, which can vary significantly between cities. For example, the City of Providence requires each operator to have at least 5\% of trips occur in each of five designated regions, shown in Figure~\ref{fig:service} \citep{provMicro}. These regulations reflect the city's priority to have consistent and widespread service and to ensure that the multiple operators that serve Providence don't just serve those neighborhoods with very high demand and low idle times. 

Because availability affects demand, equitable access correlates with the actual underlying demand when dependable and nearby access exists. To be equitable, the city wants operators to provide high service level for all areas with significant user demand. This underlying demand naturally varies across the service area given the underlying population (e.g. number of commuters or students) and built environment (e.g. distance to public transportation, access to bike lanes) \citep{chen2020dockless}. Therefore, it is essential to adjust for censoring due to unavailability when estimating demand from past usage --- the observed data will not contain events for users who wanted to use the system but did not find an available bike or scooter. Failing to consider censoring can also result in substantial overestimations of service levels \citep{albinski2018performance}. 

Expanding on the docked system literature, we present a new framework for estimating spatial-temporal demand by modeling the system as a spatial non-homogeneous Poisson process. While the docked literature has historically treated stations as independent Poisson processes with no substitution behavior \citep{freund2019bike}, we directly model the relationship between a user's initial location and where/whether a trip starts by incorporating a user choice model. This allows our model to extend to both docked and dockless systems and removes the assumption of independence between locations. We then use an Expectation-Maximization (EM) algorithm to estimate the underlying parameters from past trip and availability data. In our presentation and implementation, we incorporate a specific user choice model based on the distance between an arriving user and the available scooters or bikes, but the algorithm extends to other choice models. 

Our publicly available application implements our estimation algorithm, allowing users to analyze their data efficiently (approximately 2 minutes to analyze data with 100k trips) and was designed with feedback and input from city planners with the City of Providence to increase the usability. Additionally, because the output from the algorithm is a matrix of estimated arrival rates by location and time, the application visualizes this distribution for the user and highlights areas with potential unmet demand. By focusing on the usefulness and ease of interpretation, we aim to increase the audience for these tools and allow more people without data analysis training to gain insight from publicly available data.  

In Section~\ref{sec:lit_rev}, we place our work in the context of past literature on estimating demand in shared micromobility networks. In Section~\ref{sec:data}, we introduce the data collected from the Providence, R.I. systems and highlight the key considerations for these stakeholders. We then introduce our general framework and user choice model in Section~\ref{sec:model} and present our estimation algorithm in Section~\ref{sec:est}. In Section~\ref{sec:comp}, we present computational experiments comparing our algorithm to previous methods and perform a sensitivity analysis. Last, in Section~\ref{sec:ex}, we demonstrate the use of our publicly available tool using data from Providence, Rhode Island and comment on the results. Additionally, a small test data set, modified from Kansas City's Microtransit network \citep{kansasMicro}, is available for users to test our application. 
While our tool gives valuable insight into usage patterns and highlights potential areas to increase service levels, future work will focus on using the results from this estimation to analyze how changing access explicitly affects user behavior and overall observed demand. We discuss these possible extensions in Section~\ref{sec:future}.

\section{Literature Review} \label{sec:lit_rev}

Early analysis of bike-share data primarily concentrated on estimating demand within docked systems characterized by fixed stations. In these scenarios, station-specific demand was frequently modeled as a non-homogeneous Poisson arrival process, where the arrival rate at a particular station during a given time period was determined by dividing the observed number of trips during that time period by the proportion of time in which at least one bike was available \citep{freund2019analytics, freund2019bike}. This adjustment was essential to account for potential censoring in the data. In a related study, \cite{gammelli2020estimating} introduced a censored likelihood function within a Gaussian process model. However, these results overlook the spatial interdependence among stations, even when stations were situated in close proximity. When one station had no available bikes, riders may opt for a nearby station. Our work also uses a non-homogeneous Poisson arrival process to model user arrivals but incorporates a user choice model to model these substitution patterns. Additionally, our model is flexible enough to work for either docked or dockless systems. 

A separate strand of research has concentrated on leveraging the estimated Poisson process model for docked systems to inform operational and planning decisions. Extensive studies have focused on optimizing station inventory levels \citep{raviv2013optimal}, rebalancing of stations \citep{raviv2013static, chemla2013bike, nair2013large, dell2014bike, erdougan2014static, ho2014solving, forma20153, liu2016rebalancing, schuijbroek2017inventory, chung2018bike, datner2019setting}, and determining the optimal number of docks at stations \citep{freund2017minimizing}. For an overview of how demand models inform planning, consult related survey papers \citep{freund2019analytics, freund2019bike, shui2020review}. By also using a Poisson process model for demand, our estimation algorithm can be used in conjunction with these other operational planning results. 

A few papers look at user choice models outside of demand estimation. To investigate how distance influenced station preferences, \cite{kabra2020bike} employed a structural demand model rooted in a random utility maximization framework that considered distance to the nearest station. Their model revealed an exponential decline in the likelihood of selecting a station as the distance from it increased and estimated that approximately 80\% of bike-share usage occurred within 300 meters of stations. Similarly, \citep{he2021customer} utilized a structural demand model to estimate demand for start-end station pairs based on distance, emphasizing the importance of understanding demand when contemplating network expansion. Our chosen user choice model incorporates a similar dependence on distance as these structural demand models but within a dockless context.

In the realm of docked systems, another line of research has centered on trip rather than demand prediction, where trips are realized demand. These methods do not account for availability and have ranged from regression-based methods \citep{de2015estimating, rudloff2014modeling, zhang2016bicycle} to simulation-based methods \citep{singhvi2015predicting}. To account for spatial dependence, some of these methods have employed clustering to define neighborhoods for predicting trip volumes or destinations \citep{dai2018cluster, feng2018hierarchical}. More recently, machine learning methods such as gradient-boosted regression trees \citep{li2015traffic} and attention-based networks \citep{pan2019predicting, wang2020attention} have been applied to enhance trip prediction accuracy within docked systems. While useful for understanding the number of trips observed, these methods do not return demand estimates and are not as useful for operational planning. 

As mentioned previously, dockless systems introduce additional user flexibility and necessitate distinct methods for predicting demand and usage. Unlike docked systems, dockless systems allow operators to adjust availability without altering station infrastructure, potentially leading to increased equity \citep{mooney2019freedom, chen2020dockless}. However, failing to consider censoring can result in substantial overestimations of service levels due to the wider availability of bikes \citep{albinski2018performance}. 

To predict trips or demand in dockless systems, a spatial-temporal representation of the system is crucial, often represented as a grid or graph network. Models predicting trips have incorporated this structural aspect and employed various machine learning methods, such as random forests \citep{yang2016mobility}, gradient boosted decision trees \citep{lin2020revealing}, convolutional neural networks \citep{xiao2021demand}, and long short-term memory neural networks \citep{xu2018station, li2019learning}. While these complex methods enhance prediction accuracy, they may yield less interpretable outcomes for city planners. \cite{gu2020exploiting} attempted to identify a concise summary representation of travel between different regions within the service area, aiming to provide more digestible insights. Notably, there has been limited research on directly estimating demand rather than trips in dockless systems. To our knowledge, the only previous work to estimate demand in dockless systems has first transformed the system to a docked one by treating each grid point as an independent `station' \citep{albinski2018performance}. Again, this ignores the spatial dependence between grid cells.

Our approach is the first to focus on estimating demand for dockless systems and builds on the docked literature by removing the assumption of independence between locations and instead incorporating a user choice model. This bridges the gap between the docked and dockless literature while still allowing our model to be capitalize on the related operational planning research. Additionally, our methodology is publicly available and implemented in a user-friendly application. 

\section{Background} \label{sec:data}

\begin{figure}[h]
\begin{center}
\caption{The average number of trips observed in Providence in 15 minute intervals during the available service time, 6am to 10pm, with error bars.} \label{fig:providence_trips_data}
\includegraphics[scale=0.7]{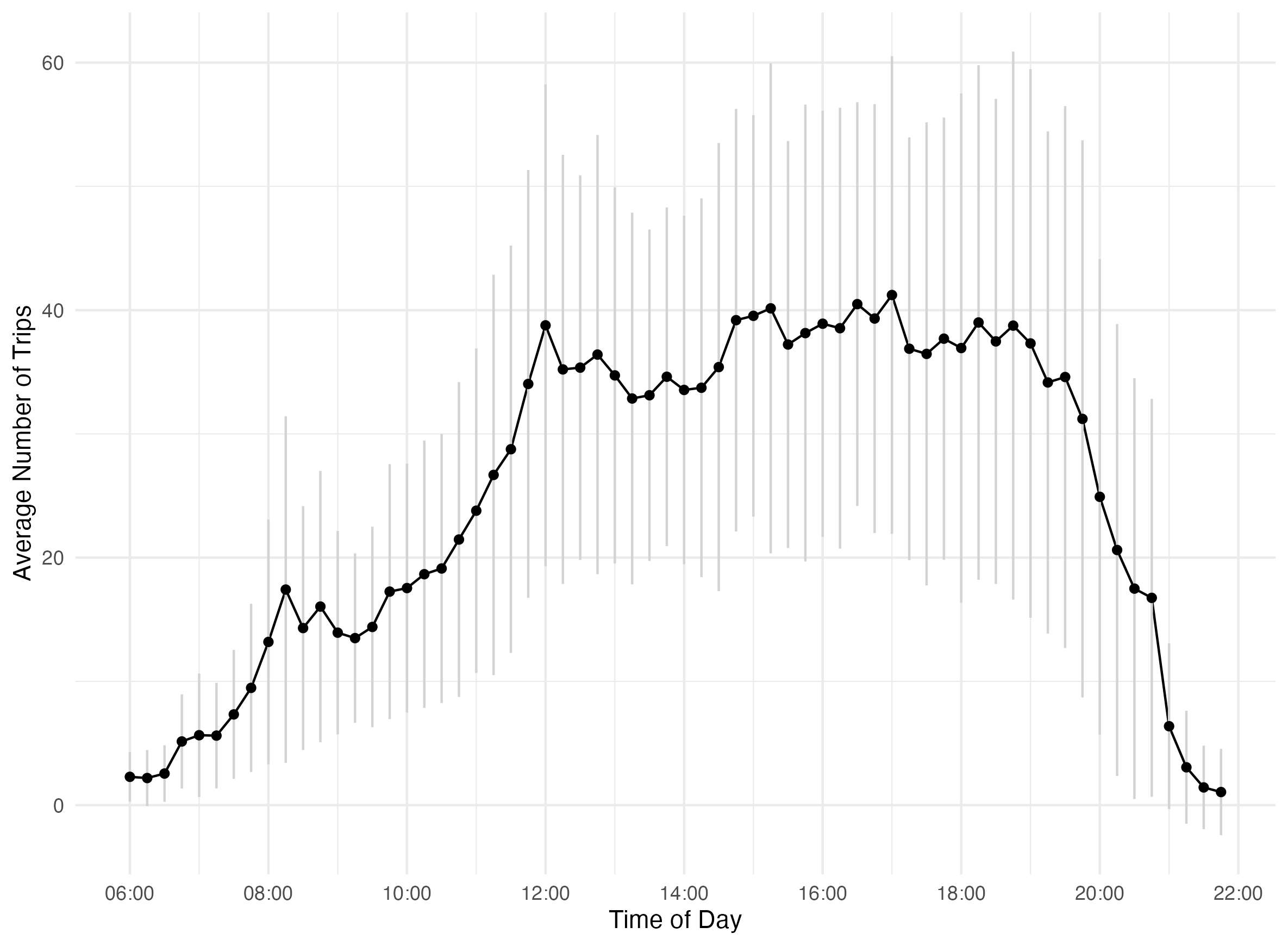}
\end{center}
\end{figure}

This research was conducted in collaboration with city planners from the City of Providence, including Principal Planner Alex Ellis and Curbside Administrator Liza Farr. Our study focuses on the bike and scooter share programs in Providence, Rhode Island, which were initiated in the fall of 2018. The data presented in this article pertains to scooter usage in Providence from June 1, 2019, to August 31, 2019, during which we observed consistent ridership. During this period, the City of Providence interacted with the system data using Remix, a transportation planning platform designed for handling General Transit Feed Specification (GTFS) data \citep{remix}. Figure~\ref{fig:providence_trips_data} illustrates the average number of trips observed over time across the city during this time period.

Notably, the city currently requires that 5\% of all trips occur in specific regions, as depicted in Figure~\ref{fig:service}. However, operators frequently fell short of meeting this requirement. The city sought to determine whether this was due to a shortage of scooter availability or lack of demand, prompting a reevaluation of the designated regions. For instance, while there was consistently high scooter availability around Brown University's campus in the `East' region, the `South' region exhibited greater variability in scooter availability, which could impact the observed trip numbers.

Through these discussions, it became evident that accurately estimating demand as opposed to recorded trips was a crucial initial step. This required an expansion of existing methodologies and the development of a model that could be used in future urban planning decisions. We chose to model demand as a stochastic process, employing a non-homogeneous Poisson process coupled with a user choice model. This approach builds upon established methodologies in the literature related to docked transportation systems and enables future planning decisions to leverage insights from operations planning literature. Moreover, our model provides interpretable outputs in the form of arrival rates by location and time and our application highlights regions with low service levels. Our chosen user choice model was also guided by discussions with the city and can be easily defined by two interpretable parameters. Furthermore, our methodology was designed to facilitate the city's utilization through a user-friendly web application, allowing stakeholders to visualize and download the results with ease.

To summarize, this research focused on the following key aims raised by the city authorities:

\begin{enumerate}
    \item Estimation of the underlying demand across the city.
    \item Assessment of the equity of scooter availability in relation to demand by identifying areas with low service levels but significant demand.
    \item Planning for the potential utilization of these findings to inform urban planning decisions, including potential updates to contracts with the service operators and adjustments to the total number of available scooters.
\end{enumerate}

\section{Framework} \label{sec:model}

\begin{figure}[h]
\begin{center}
\caption{Visualization of blank grid discretization (400m grid cells, Providence).} \label{fig:blank_grid}
\includegraphics[scale=0.5]{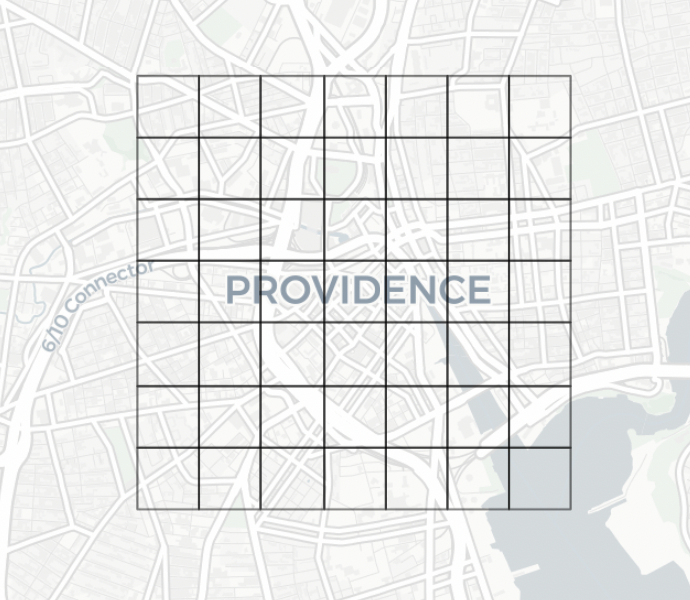}  
\end{center}
\end{figure}

This section defines our general framework and notation along with our chosen user choice model for how users arrive and make decisions. For ease of presentation, we consider our shared micromobility network to only have available bikes (as opposed to scooters or both bikes and scooters). We first discretize the space by inducing a grid over the network and indexing the grid cells $\mathcal{G} = \{1, 2, \ldots, m$\}. See Figure~\ref{fig:blank_grid} for an example of a 400m grid induced over Providence. Further, we discretize the times of day into discrete time periods $h \in \{ 1, 2, \ldots, \mathcal{H}\}$. In our app, these time periods are set to the hours of the day and the user specifies the granularity of this grid — smaller grid cells give more detailed demand estimates but require more computation time.

Our framework relies on the following three assumptions.
\begin{enumerate}
    \item All bikes and users are at the center point of the corresponding grid cell. 
    
    \item Users arrive in the center of each grid cell $i \in \mathcal{G}$ following an independent non-homogeneous stochastic Poisson process where the rate parameter $\mu_{h,i}$ depends on the time period $h \in \{ 1, 2, \ldots, \mathcal{H}\}$.
    
    \item Each user arriving in grid cell $i$ observes the distribution of available bikes and either picks a bike from grid cell $j \in \{1, 2, \ldots, m\}$ with probability $\mathrm{prob}_{j}$ or leaves the system without using a bike with probability $\mathrm{prob}_{0}$, where $ \mathrm{prob}_{0} + \sum_{j=1}^m \mathrm{prob}_{j} = 1$. The probabilities $\mathrm{prob}_{j}$ incorporate the censoring and spatial dependence between a user's location and any generated trip. 
\end{enumerate}

The framework above assumes that the rate $\mu_{h,i}$ is fixed for each time period and grid cell. That is, the estimated stochastic Poisson process can account for the variance in demand. In Section~\ref{subsec:ext}, we discuss the dependency on other covariates such as the day of week, weather, season or other characteristics. We expect that incorporating weather and day of the week in particular will improve the model accuracy. 
Note that our framework and estimation algorithm are agnostic to the distribution of the probabilities $\mathrm{prob}_j$ and can be generalized to any choice model. For example, a user choice model may consider the direction of travel.

\subsection{User Choice Model}

\begin{figure}
\begin{center}
\caption{In the 400m grid, when a user arrives in the center cell with a distance threshold of at least 400m, they randomly choose one of the 3 bikes that are 400m away in the closest neighboring cells. The user would never consider the bike that is 800m to the left.} \label{fig:user_behavior}
\vspace*{0.3cm}
\includegraphics[scale=0.35]{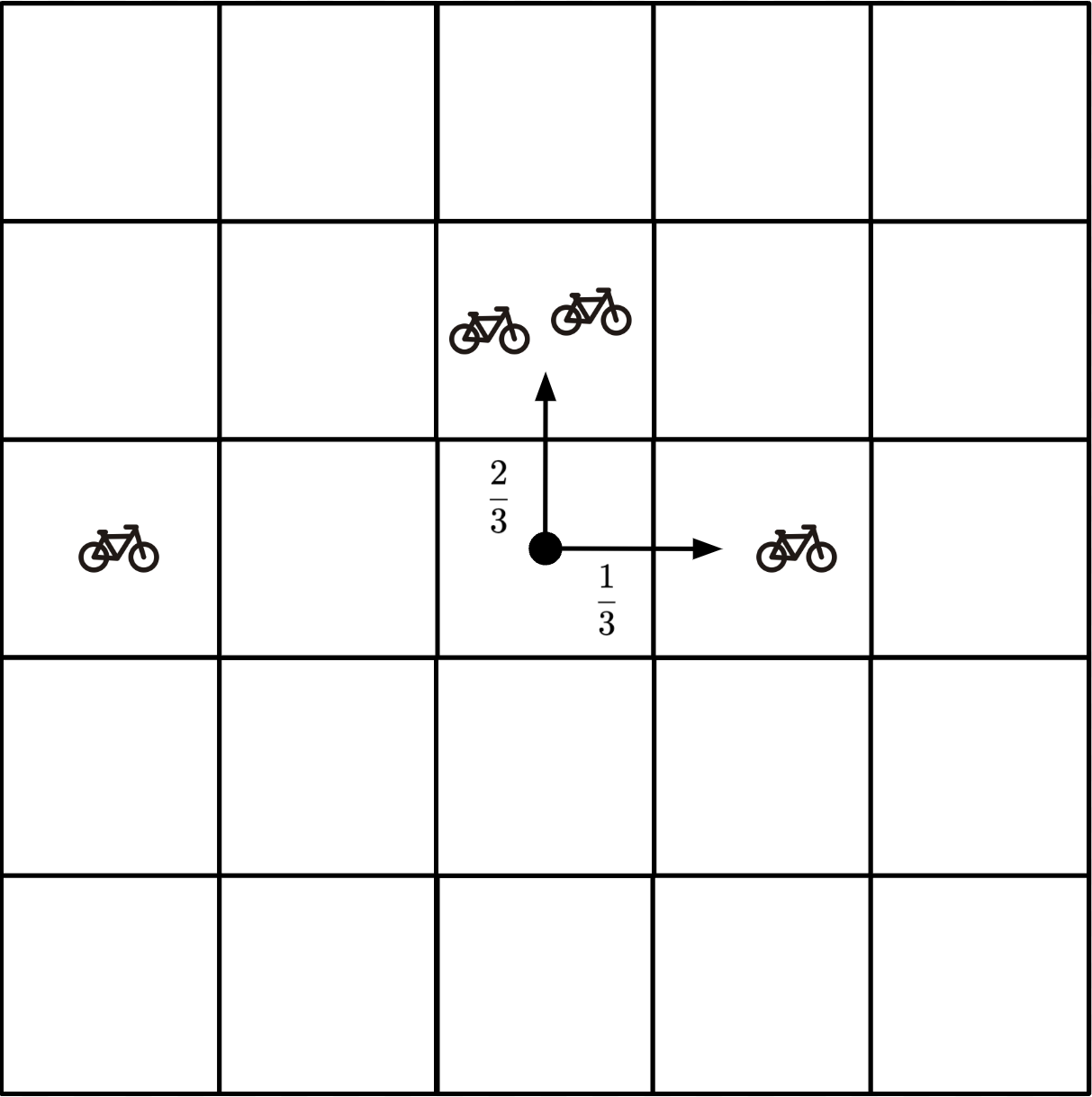}
\end{center}
\end{figure}

We introduce a choice model in which a user's choice only depends on the distance traveled. We assume arriving users have a threshold distance they are willing to travel to pick up a bike. Users then greedily choose an available bike within that threshold, breaking ties randomly. If no bike is within the user's threshold, they leave without generating a trip. For example, in the 400m grid in Figure~\ref{fig:user_behavior}, if a user arrives in the central cell with a distance threshold of at least 400m, they randomly choose one of the 3 bikes that are 400m away in the closest neighboring cells. The user would never consider the bike that is 800m to the left since users greedily choose one of the closest available bikes. If the user's distance threshold is less than 400m, the user would leave without generating a trip.

\begin{figure}
\begin{center}
\caption{Visualization of relative grid cell distances in a 400m grid with a maximum distance of 1000m.} \label{fig:grid_dist}
\includegraphics[scale=0.15]{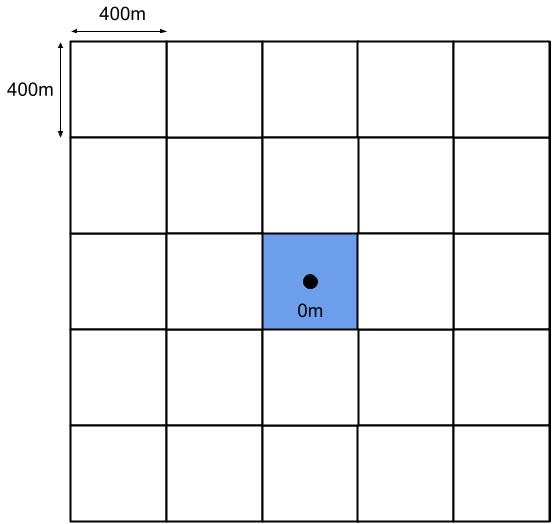}  
\includegraphics[scale=0.15]{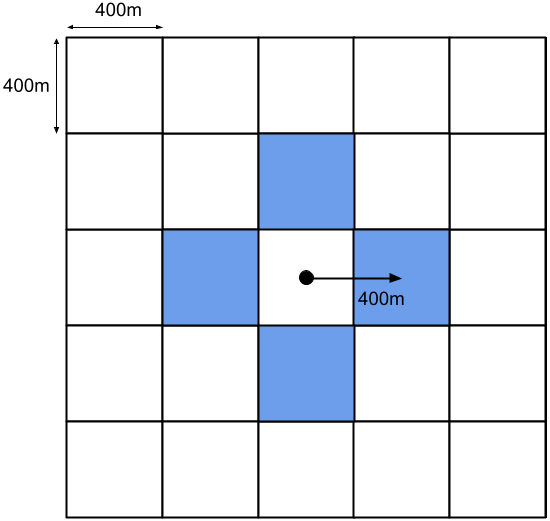}  
\includegraphics[scale=0.15]{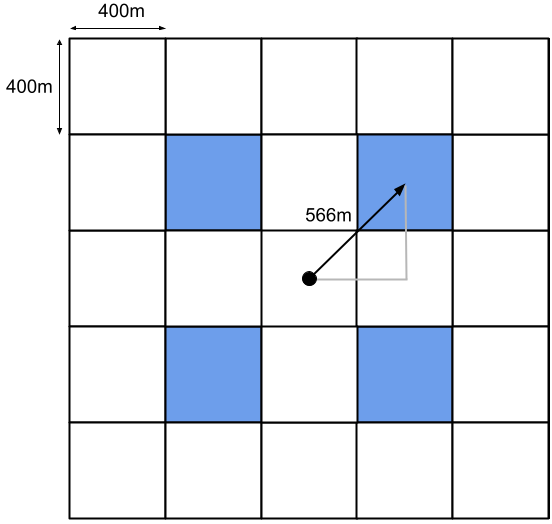}  
\includegraphics[scale=0.15]{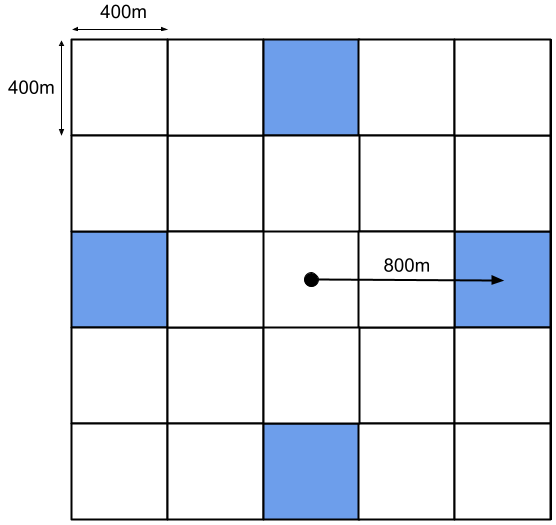}
\includegraphics[scale=0.15]{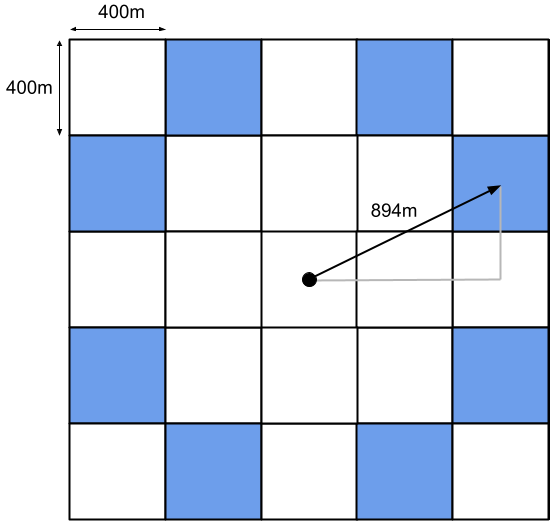}
\end{center}
\end{figure}

\begin{figure}
\begin{center}
\caption{On the left, the truncated probability density function for the folded half normal distribution with variance $\mathbf{\sigma^2}$ is given by the blue curve and the corresponding user threshold distribution is given by the bar plot. On the right, example of how $\mathbf{p_0}$ affects the estimated distribution.} \label{fig:halfnorm}
\includegraphics[scale=0.35]{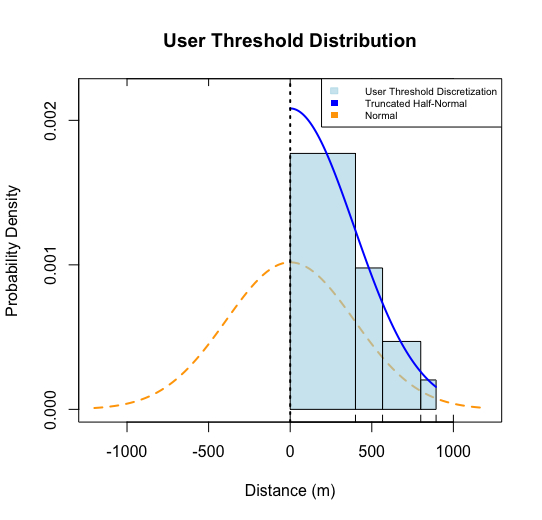}
\includegraphics[scale=0.35]{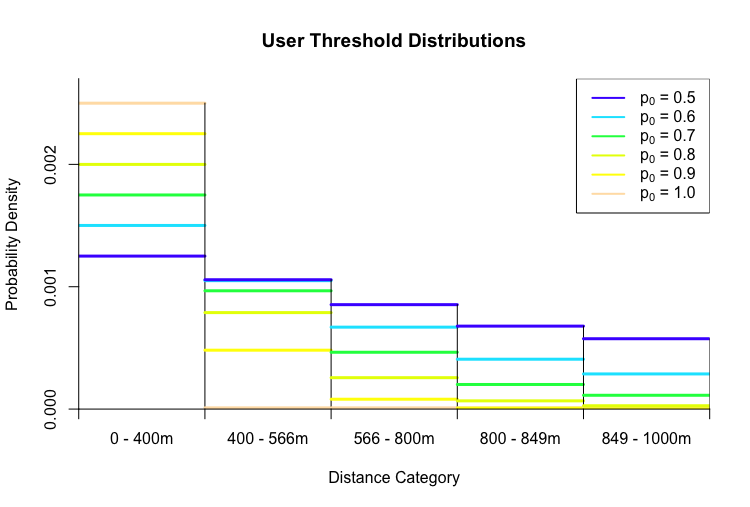}
\end{center}
\end{figure}

The distribution of user thresholds is informed by our conversations with city planners and previous work on user demand models \citep{he2021customer, kabra2020bike}. In particular, we use a discretized version of a half-normal distribution as shown in Figure~\ref{fig:halfnorm}. This distribution allows us to control the decay in the likelihood of user travel as distance increases. For a half-normal distribution with standard deviation $\sigma$, let $\mathrm{f}_{\sigma}(\cdot)$ be the corresponding cumulative distribution function. We consider all possible distances $\mathcal{D}=\{\mathrm{dist}_0, \mathrm{dist}_1, \ldots, \mathrm{dist}_{\text{max}} \}$ between the center points of two grid cells up to distance $\mathrm{dist}_{\text{max}}$, the maximum distance any user would be willing to walk to find a bike and define for $l \in \{0,1, \ldots \text{max}-1 \}$ the probability a user has a threshold in the range $[\mathrm{dist}_l, \mathrm{dist}_{l+1})$ to be 
\[ \mathrm{Pr}(\mathrm{dist}_{l} \leq \text{threshold} < \mathrm{dist}_{l+1} ) := (f_{\sigma}(\mathrm{dist}_{l+1}) - f_{\sigma}(\mathrm{dist}_{l}))/(1-f_{\sigma}(\mathrm{dist}_{\text{max}})).\] 
The division by $(1-f_{\sigma}(\mathrm{dist}_{\text{max}}))$ ensures the probabilities sum to one. Given the assumption that all bikes and users are at the center point of a grid cell, this is the probability a user considers bikes up to a distance $\mathrm{dist}_l$ away. Figure~\ref{fig:grid_dist} shows how the grid induces the set of possible travel distances, and Figure~\ref{fig:halfnorm} demonstrates how the folded normal distribution leads to the discretized threshold probabilities for a set maximum distance and $\sigma$.

To set $\sigma$, we let $p_0$ be the probability a user is only willing to consider bikes within their grid cell, a more interpretable parameter, and use bisection search to find $\sigma$ such that $\mathrm{Pr}(\mathrm{dist}_{0} \leq \text{threshold} < \mathrm{dist}_{1} ) \approx p_0$. Specifying a higher $p_0$ reduces $\sigma$ and leads to a sharper decrease in probabilities, whereas a small $p_0$ can lead to an almost uniform distribution on threshold up to the maximum distance, showing the flexibility in this distribution. Figure~\ref{fig:halfnorm} illustrates this for different values of $p_0$. 
Past research on user behavior indicates that a maximum travel distance of 800m or 1000m is reasonable and that there is likely an exponential decay in the likelihood of travel as distance increases \citep{zhang2019best,eren2020review, he2021customer, kabra2020bike}. In practice, city planners could create user surveys to determine good estimates for the maximum travel distance and $p_0$. In our public application, $\mathrm{dist}_{\text{max}}$ and $p_0$ are user-specified parameters with default values of 1000m and 0.7, respectively, for a grid with 400m wide grid cells, based on discussions with the City of Providence. Users can re-run the algorithm to analyze how demand estimates change with these parameters or to get a range of demand estimates.  

\section{Estimation Algorithm} \label{sec:est}

To estimate the Poisson arrival rates $\mu_{h,i}$, we assume we have data on the observed trips and the location of available bikes across $k$ days. Suppose there are $n$ observed trips and that each trip $x \in \{ 1,2, \ldots, n \}$ occurs at time $t_x$ in time period $h_x$ in grid cell $j_x$. We let the tuple $y_x = (t_x, h_x, j_x)$ represent each trip. Our goal is to set $\mu_{h,i}$ to maximize the likelihood of the observed data. 

As mentioned, estimating the rates is complicated by the censoring of a user's starting location and potentially unobserved demand. Users may walk from an unobserved cell location to an available bike masking the true location of user demand, and users may look for a bike and not find one within their user threshold, which is unobserved user demand. To address these limitations, we define an indicator latent variable $z_{x,i}$ representing whether the user from trip $x$ arrived from cell $i$ before choosing a bike in cell $j_x$ and use an Expectation-Maximization algorithm to infer the Poisson arrival rates of each cell during each time period. 

\subsection{Algorithm Notation} \label{subsec:alg_not}

Before defining the algorithm, we define two probabilities necessary for the calculations. First, we define $\pi_{x,i}$ to be the probability a user arriving in cell $i$ at the time of trip $x$ would choose a bike in cell $j_x$, the cell from which trip $x$ begins. These probabilities allow us to infer the likelihood that an observed trip in cell $j_x$ arrived from cell $i$ when estimating the arrival rates. Second, we define $\alpha_{h,i}$ to be the probability a random user arriving in cell $i$ will find a bike within their distance threshold. This value estimates the proportion of arriving users from cell $i$ who are unobserved in order to accurately estimate the arrival rates. 

We define $\pi_{x,i}$ to be the probability a user arriving in cell $i$ at the time of trip $x$ would choose a bike in cell $j_x$, the cell from which trip $x$ begins. Let $\mathrm{dist}_{x,i}$ be the distance between grid cell $i$ and cell $j_x$, $S^1_{x, i}$ be the set of available bikes closest to grid cell $i$ at the time $t_x$, and $S^2_{x,i}$ be the set of bikes in $S^{1}_{x,i}$ within grid cell $j_x$. Note that $S^{1}_{x,i}$ contains all bikes in the grid cells closest to cell $i$ among those with at least one bike. Then the probability a user in grid cell $i$ would choose a bike in grid cell $j_x$ is based on whether the user's threshold is at least $\mathrm{dist}_{x,i}$ and the fraction of bikes in $S^1_{x,i}$ that are in grid cell $j_x$. 
    \[\pi_{x,i} =  \text{Pr}(\text{threshold} \geq \mathrm{dist}_{x,i}) \cdot \frac{|S^2_{x,i}|}{|S^1_{x,i}|}.\]
    Note that when $j_x$ is further from cell $i$ than another cell that contains bikes, $S^2_{x,i} = \emptyset$ and this probability is zero.


Second, we define $\alpha_{h,i}$ to be the probability a user arriving in cell $i$ at a uniformly distributed time within time period $h$ and with a user threshold following the specified user threshold distribution will find a bike within their distance threshold. To estimate this probability, we consider each possible threshold value range. Suppose that the user's threshold is in the range $[\mathrm{dist}_l, \mathrm{dist}_{l+1})$ and consider the availability during time period $h$. Using the bike locations during that time period, we can find the percent of time that the closest bike to cell $i$ is at most relative cell distance $\mathrm{dist}_l$ away from cell $i$.  Let this value be $\mathrm{perc}_{h,i,l}$. This is calculated by keeping track of the distribution of the closest bike over time and finding the average number of minutes the closest bike is at most distance $\mathrm{dist}_l$. 

Then, to find $\alpha_{h,i}$, we consider all possible distances for the closest bike. Summing these probabilities over all feasible distances and then averaging across the days, we obtain an empirical estimate for $\alpha_{h,i}$.
\[ \hat{\alpha}_{h,i} = \sum_{l=0}^{|\mathcal{D}|-1} \mathrm{perc}_{h,i,l} \cdot \mathrm{Pr}(\mathrm{dist}_{l} \leq \text{threshold} < \mathrm{dist}_{l+1} ) .\]


\subsection{EM Algorithm} \label{subsec:em_alg}

As mentioned earlier, we introduce the indicator latent variable $z_{x,i}$ representing whether the user from trip $x$ comes from cell $i$. Since each trip can come from only one cell, we have the constraint that $\sum_{i=1}^m z_{x,i} = 1$. We now consider the full data $\{(y_x,z_x) \text{ for } x = 1, 2, \ldots n \}$ where $y$ is the observed trip data and $z$ is the unobserved data. The log-likelihood of the observed data with arrival rates $\mu$ can be written as
\begin{align*}
    \ell(\mu ; x) &=  \sum_{x=1}^n \log p(y_x| \mu) \\
    &= \sum_{x=1}^n \log \sum_{i=1}^m  p(y_x, z_{x,i} = 1 | \mu)  
\end{align*}
We use an EM algorithm to maximize the log-likelihood function by alternating between two steps: an Expectation Step (E-Step) and a Maximization Step (M-Step). In the E-Step, we maximize the expectation of log-likelihood with respect to the latent indicator variables $z$ given the trip data $y$ and current estimates for $\mu$, and in the M-Step, we maximize the log-likelihood estimate with respect to the parameters $\mu$ given the trip data $y$ and the current probability estimates for $z$. The algorithm is guaranteed to converge to a local optima \citep{mclachlan2007algorithm}.

In the expectation step of the algorithm, we maximize the log-likelihood with respect to the distribution of $z$ given trip data and $\mu$. Since \[p(y_x, z_{x,i} = 1 | \mu) = p(z_{x,i} = 1 | y_x, \mu) \cdot p(y_x| \mu)\] we can maximize the log-likelihood by estimating the first term, the posterior probability that $z_{x,i} = 1$ given the data and arrival rates $\mu$. Now $p(z_{x,i} = 1 | y_x, \mu)$ is the probability that the user from trip $x$ at time $t_x$ arrived in cell $i$ and chose a bike in cell $j_x$. We'll define this likelihood to be the membership weight of trip $x$ to cell $i$ $w_{x,i}$, which can be derived as follows.

Given a mixture of $m$ independent Poisson processes with average rates $\mu_{h_x,1}, \mu_{h_x,2}, \ldots, \mu_{h_x,m}$ during time period $h_x$, the probability that a user is from Poisson process $i$ is given by $\mu_{h_x,i}/\sum_{l=1}^m \mu_{h_x,l}$. Using the probability $\pi_{x,i}$ that an arriving user in cell $i$ at the time of the trip $x$ chooses the bike from trip $x$ in cell $j_x$, we find that the overall probability is given by
\begin{align*}
   \pi_{x,i} \cdot \frac{\mu_{h_x,i} }{ \sum_{l=1}^m  \mu_{h_x,l}}.
\end{align*}
We then normalize this probability distribution over all cells to obtain the likelihood membership weights $w_{x,i}$.
\begin{equation}
\label{eqn:mem_weights}
    w_{x,i}  
    = \frac{\left( \frac{\pi_{x,i} \mu_{h_x,i} }{ \sum_{l=1}^m  \mu_{h_x,l}} \right)}{\sum_{i'=1}^m \left( \frac{\pi_{x,i'} \mu_{h_x,i'} }{ \sum_{l=1}^m  \mu_{h_x,l}}\right)} 
    = \frac{\pi_{x,i} \mu_{h_x,i} }{\sum_{i'=1}^m \pi_{x,i'} \mu_{h_x,i'}}
\end{equation}
Intuitively, the membership weights balance the arrival rates by the likelihood a user would actually travel to cell $j_x$ from cell $i$ at time $t_x$.

In the maximization step of the EM algorithm, we maximize the estimate for log-likelihood with respect to $\mu$ given the data and the current distribution of $z$ found in the E-Step. Estimating the likelihood as $p(y_x, z_{x,i} =1 | \mu) = p(z_{x,i} = 1 | y_x, \mu) \cdot p(y_x | z_{x,i} = 1, \mu) $, we get
\[ \hat{\mu} = \arg \max_{\mu} \sum_{x=1}^n \log \sum_{i=1}^m  w_{x,i} p(y_x | z_{x,i} = 1, \mu). \]

Since the arrivals in cell $i$ follow a Poisson process, given a fixed number of arrivals, the distribution of those arrivals over time period $h$ is uniformly distributed. 
Recall the probability $\alpha_{h_x,i}$ that a uniformly distributed user arriving in cell $i$ during time period $h$ finds a bike within their threshold. Then the number of observed points follows a Poisson distribution with mean $\alpha_{h_x,i} \cdot \mu_{h_x,i}$. Therefore, setting $\mu_{h_x,i}$ to maximize the likelihood of observing the trip data will yield estimate
\begin{equation}
\label{eqn:mu_update}
\hat{\mu}_{h,i} =   \frac{ \frac{1}{k}\sum_{x:h_x = h} \hat{w}_{x,i}}{\hat{\alpha}_{h,i}}, 
\end{equation}
where $\hat{w}_{x,i}$ is calculated using Equation~\ref{eqn:mem_weights} using the current estimate of $\hat{\mu}$.

In summary, we can find estimated rates $\hat{\mu}$ by alternating between the following two steps.
\begin{enumerate}
    \item{E-Step:} Maximize the expectation of log-likelihood with respect to the latent indicator variables $z$ given the trip data and current arrival rates $\hat{\mu}$ by using Equation~\ref{eqn:mem_weights} to update the estimated membership weights $\hat{w}_{x,i}$.
    \item{M-Step:} Maximize the log-likelihood estimate with respect to the parameters $\mu$ given the trip data and the current membership weights $\hat{w}_{x,i}$ by using Equation~\ref{eqn:mu_update} to update the estimated arrival rates $\hat{\mu}_{h,i}$.
\end{enumerate}
We initialize the rates $\hat{\mu}$ to be uniform across all grid cells and discuss this initialization in Secion~\ref{sec:comp}.

\subsection{Extensions} \label{subsec:ext}

We conclude this section by describing extensions to the above framework. In Section~\ref{sec:model}, we assumed Poisson arrival rates did not change day to day. Suppose we want to allow certain factors to impact arrival rates, such as the weather, season, or day of the week. In that case, we can filter the data before analysis or extend the model to estimate a rate $\mu_{h,i}$ for each day by allowing for a weighted average in the M-step, with weights reflecting the similarity between days using the specified characteristics. Additionally, we can extend the model to predict demand for trips rather than just starting locations by using the distribution of drop-off locations. Drop-off locations have the added benefit of being uncensored in the case of dockless systems. Last, the algorithm above is also flexible enough to extend beyond bike-share or scooter-share programs to any settings in which we are interested in estimating rates of events where the observed events depend on some underlying availability. For example, to model demand for bus travel routes when users may switch between buses depending on the schedule and time of day.

\section{Computational Experiments} \label{sec:comp}

In this section, we compare the performance against the estimation method given in \citep{albinski2018performance}. Our demand estimation algorithm was implemented in python 3.10.9 and the code is available at \href{https://github.com/KyFlynn/shared-mobility-research}{https://github.com/KyFlynn/shared-mobility-research}. We also analyze the sensitivity of our algorithm to the initial values of $\mu$.

\subsection{Experiment Setup} \label{subsec:comp_exp}

We define the Naive demand algorithm to be the algorithm which estimates the demand $\mu_{h,i}$ as the average rate of observed trips in grid cell $i$ in time period $h$ divided by the observed proportion of time in which a bike is available. This treats each grid cell as independent and assumes that users are not willing to travel, even when the set grid cell size is very small. In our EM algorithm, we set the initial rates uniformly so all grid cells start with the same initial guess of demand.

\begin{figure}[h]
    \centering
    \includegraphics[scale=0.3]{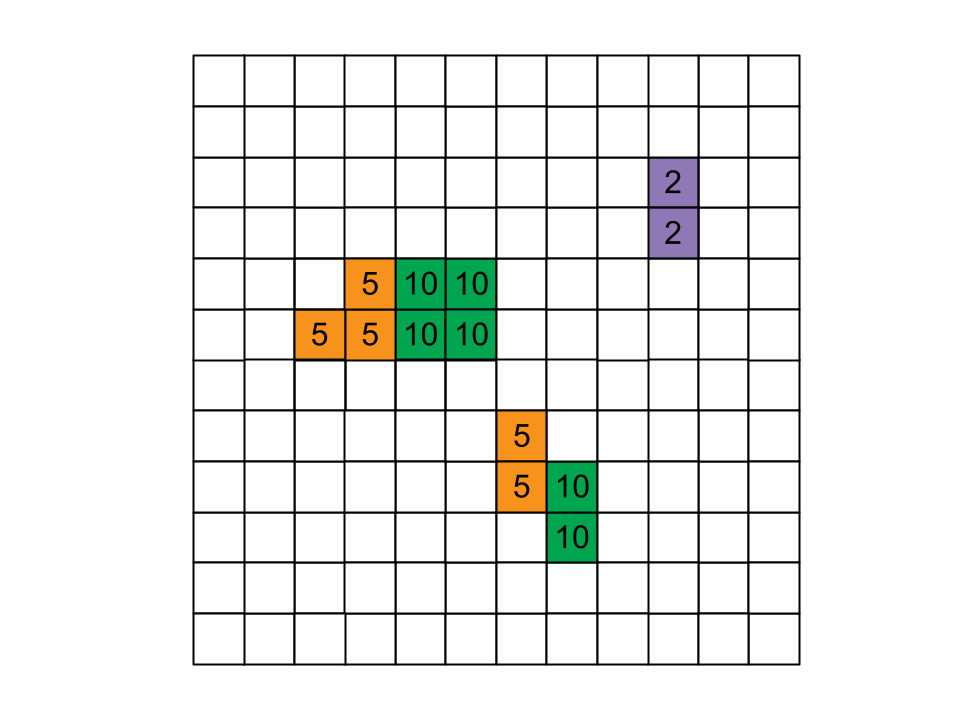}
    \caption{Visualization of the grid of demand for the sensitivity analysis data simulation. Empty cells have no demand. }
    \label{fig:sens_medium_complex_demand}
\end{figure}

To better understand the sensitivity of the demand model to surrounding availability and to compare against an estimation algorithm that does not include user choice, we conduct computational experiments in which the level of availability is varied. For each generated data set, we create a 12x12 400m grid and simulate user arrivals during a single time period $h$ over a period of 30 days where each grid cell is assigned a Poisson demand rate $\mu_{h,i}$ shown in Figure~\ref{fig:sens_medium_complex_demand}. We set these rates by defining types of grid cells to reflect what might be observed in practice, inspired by the Providence data. In particular, we consider the following four cell types.
\begin{enumerate}
    \item \textbf{\color{Green} Cluster center grid cells} have rate $\mu_{h,i} = 10$ and constant availability. These represent areas with high demand that are well-served by the operators. 
    \item \textbf{\color{BurntOrange} Border cluster cells} are cells that border the cluster centers and have a lower rate $\mu_{h,i}=5$. 
    \item \textbf{\color{RoyalPurple} Isolated demand cells} are far from the cluster centers and have a low rate of $\mu_{h,i}=2$.
    \item \textbf{No demand cells} have demand $\mu_{h,i} = 0$.
\end{enumerate}
Arriving users have a random threshold they are willing to travel, and greedily choose the nearest available bike within this threshold, breaking ties arbitrarily. The user threshold distribution is set with a maximum distance of 1000m and $p_0=0.7$. Experiments with $p_0 = 0.5$ were also tested and yielded similar results, with the EM algorithm further outperforming the Naive algorithm since users are more willing to travel. 

When there is perfect availability, there is no substitution behavior or lost users and the two algorithms return the same demand estimates. In practice, areas with the highest demand have very high availability but other areas may have sporadic availability. To model this, we assume that cluster centers have perfect availability with an infinite supply of bikes. For all other grid cells, for each day there is a probability $p$ that there is perfect availability and with probability $1-p$ there are no bikes available in that cell that day. For each value of $p \in \{0, 0.1, \ldots, 0.9, 1.0\}$ we generate 10 data sets where each data set records the trips and availability in time period $h$ for 30 days. 

\subsection{Results}
\label{subsec:comp_results}

\begin{figure}
     \centering
        \caption{Median and maximum error by cell type for our Expectation-Maximization (EM) algorithm and the Naive estimation algorithm for varying availability. The gray line shows the median and maximum error for the realized average user arrival rate for comparison. }
        \label{fig:comp_results}
     \begin{subfigure}[b]{0.45\textwidth}
         \centering
         \includegraphics[width=\textwidth]{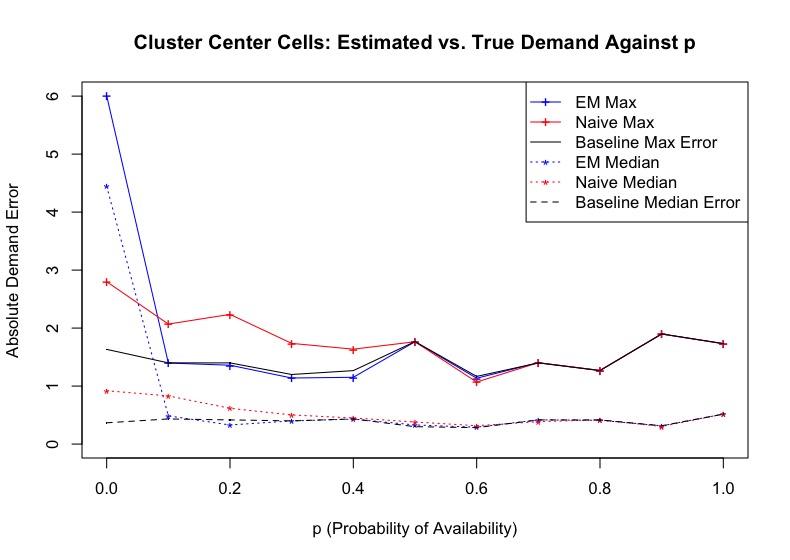}
         \caption{Cluster Center Cells}
         \label{fig:comp_results_centers}
     \end{subfigure}
     \hfill
     \begin{subfigure}[b]{0.45\textwidth}
         \centering
         \includegraphics[width=\textwidth]{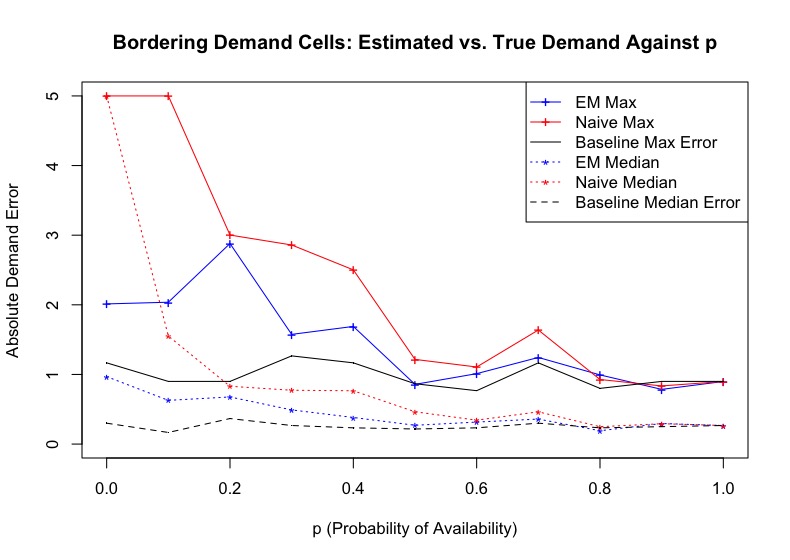}
         \caption{Bordering Demand Cells}
         \label{fig:comp_results_bordering}
     \end{subfigure}
     \hfill
     \begin{subfigure}[b]{0.45\textwidth}
         \centering
         \includegraphics[width=\textwidth]{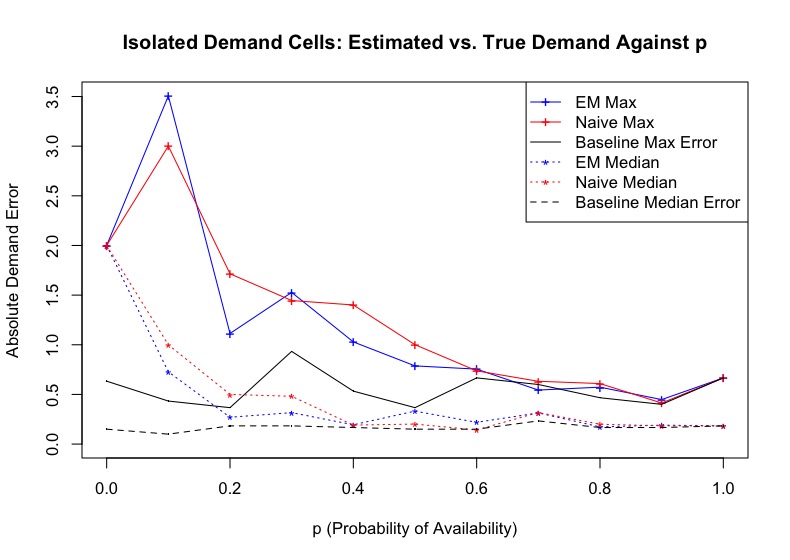}
         \caption{Isolated Demand Cells}
         \label{fig:comp_results_isolated}
     \end{subfigure}
     \hfill
     \begin{subfigure}[b]{0.45\textwidth}
         \centering
         \includegraphics[width=\textwidth]{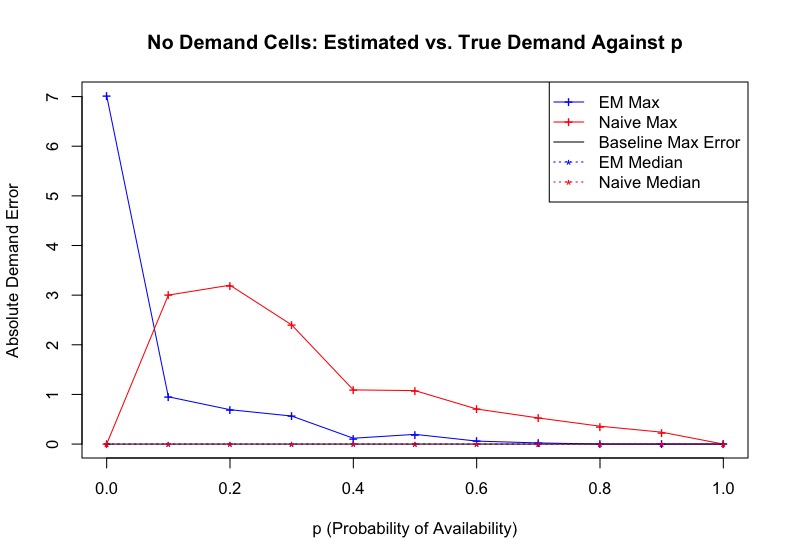}
         \caption{No Demand Cells}
         \label{fig:comp_results_no}
     \end{subfigure}

\end{figure}

Table~\ref{tab:comp_results} and Figure~\ref{fig:comp_results} show the median and maximum error for the EM (blue) and Naive (red) algorithms for each value of $p$ stratified by cell type. Additionally, in Figure~\ref{fig:comp_results}, we show the median and maximum error in the case when we have full uncensored data and could observe the true arrival rate for each grid cell (gray). For $p=0$, there is no availability outside the cluster centers and both algorithms have the highest error. In this case, the initialization of the EM algorithm spreads out demand leading to underestimation of demand for the the cluster centers and overestimation for the no demand cells. However, the EM algorithm has better estimates for the bordering demand cells. For $p$ between $0.1$ and $0.5$, however, the EM algorithm consistently outperforms the Naive algorithm for all cell types. As $p$ increases to 1, the median and maximum error for both algorithms decreases and the algorithms have similar performance as users are more likely to find a bike in their own grid cell. 

Overall, the results show in the setting with incomplete but non-zero availability the EM algorithm is able to figure out the substitution patterns of users and better estimate demand compared to the Naive algorithm. For city planners, this means that in areas with at least partial availability, the EM algorithm is better able to estimate demand, and in areas where operators have not provided any availability, there isn't enough data to estimate rates accurately. 

\begin{table}
\caption{Median and maximum error by cell type for our Expectation-Maximization (EM) algorithm and the Naive estimation algorithm for varying availability. The lowest error between the two algorithms is given in blue (EM) or red (Naive). }
\label{tab:comp_results}
\setlength{\tabcolsep}{4pt}
\begin{center}
{\small
    \begin{tabular}{|c|c|c|c|c|c|c|c|c|c|c|c|c|c|}
    \hline
    Cell & Error & & \multicolumn{11}{c|}{Probability $p$ of Availability for Non-Clusters} \\ \cline{4-14}
     Category & Measure & Algorithm &  0.0 &  0.1 & 0.2 & 0.3 & 0.4 & 0.5 & 0.6 & 0.7 & 0.8 & 0.9 & 1.0 \\ 
     \hline
     All & Median & EM & 0.86 & {\color{blue}0.00} & {\color{blue}0.00} & {\color{blue}0.00} & {\color{blue}0.00} & {\color{blue}0.00} & {\color{blue}0.00} & {\color{blue}0.00} & {\color{blue}0.00} & {\color{blue}0.00} & {\color{blue}0.00}  \\
     &  & Naive & {\color{red}0.00} & {\color{red}0.00} & {\color{red}0.00} & {\color{red}0.00} & {\color{red}0.00} & {\color{red}0.00} & {\color{red}0.00} & {\color{red}0.00} & {\color{red}0.00} & {\color{red}0.00} & {\color{red}0.00}  \\ 
     \cline{2-14}
     & Max & EM & 7.25 & {\color{blue}4.19} & {\color{blue}2.57} & {\color{blue}3.02} & {\color{blue}2.19} & {\color{blue}1.90} & 1.74 & {\color{blue}1.84} & {\color{blue}1.60} & {\color{blue}1.36} & {\color{blue}1.50}  \\
     &  & Naive & {\color{red}5.00} & 7.00 & 3.67 & 4.20 & 2.45 & 2.44 &  {\color{red}1.50} &  {\color{red}1.84} &  {\color{red}1.60} & 1.44 &  {\color{red}1.50} \\
     \hline
     Cluster  & Median &  EM & 4.42  & {\color{blue}0.46} & {\color{blue}0.39} & {\color{blue}0.35} & {\color{blue}0.40} & {\color{blue}0.37} & 0.38 & {\color{blue}0.35} & 0.40 & {\color{blue}0.45} & {\color{blue}0.42}\\
     Centers &  & Naive & {\color{red}0.94} & 0.65 & 0.45 & 0.50 & 0.57 & 0.42 & {\color{red}0.37} & 0.37 & 0.37 & 0.47 & {\color{red}0.42} \\
      \cline{2-14}
        & Max  &  EM & 6.08 & {\color{blue}1.93} & {\color{blue}1.70} & {\color{blue}1.53} & {\color{blue}1.44} & {\color{blue}1.90} &  1.74 & {\color{blue}1.84} & {\color{blue}1.60} &  {\color{blue}1.36} & {\color{blue}1.50} \\
      &  & Naive & {\color{red}3.23} & 3.64 & 2.30 & 2.07 & 1.60 &  {\color{red}1.90} & {\color{red}1.50} & {\color{red}1.84} & {\color{red}1.60} &  1.44 & {\color{red}1.50} \\
     \hline
     Bordering & Median & EM & {\color{blue}1.11}  & {\color{blue}0.96} & {\color{blue}0.66} & {\color{blue}0.58} & {\color{blue}0.53} & {\color{blue}0.36} &  {\color{blue}0.36} & 0.38 & 0.33 & 0.23 & {\color{blue}0.33} \\
      Demand & & Naive & 5.00  & 1.50 & 1.00 & 1.00 & 0.65 & 0.47 & 0.39 & {\color{red}0.36} & {\color{red}0.31} & {\color{red}0.22} & {\color{red}0.33} \\
      \cline{2-14}
       & Max & EM & {\color{blue}2.25} & {\color{blue}4.19} & {\color{blue}2.57} & {\color{blue}3.02} & {\color{blue}2.19} & {\color{blue}1.52} &  1.22 & {\color{blue}1.18} &  {\color{blue}1.42} &  {\color{blue}0.90} &  {\color{blue}0.80} \\
      & & Naive & 5.00 & 7.00 & 3.67 & 4.20 & 2.45 & 2.44 & {\color{red}1.07} & 1.34 &  1.46 & 0.93 & 0.80\\
     \hline
     Isolated  & Median & EM & {\color{blue}2.00} & 0.59 & {\color{blue}0.32} & {\color{blue}0.24} & {\color{blue}0.26} & {\color{blue}0.20} & 0.21 & 0.24 & 0.23 & 0.19 & {\color{blue}0.18} \\
    Demand & & Naive & {\color{red}2.00}  & {\color{red}0.31}  & 0.50  & 0.28  & 0.34 & 0.23  & {\color{red}0.20} & {\color{red}0.23}  & {\color{red}0.22}  & {\color{red}0.16}  & {\color{red}0.18} \\
      \cline{2-14}
       & Max & EM & {\color{blue}2.00} & {\color{blue}1.97} & 1.91 & {\color{blue}0.89} & {\color{blue}0.75} &  {\color{blue}0.80} &  {\color{blue}0.65} & 0.62 & {\color{blue}0.76} &  {\color{blue}0.75} &{\color{blue}0.57} \\
    & & Naive & {\color{red}2.00} & 2.00 & {\color{red}1.50} & 1.60 & 0.93 &0.94 & 0.93 & {\color{red}0.50} & 0.83 & 0.76 & {\color{red}0.57}\\
     \hline
     No & Median & EM & {\color{blue}0.00} & {\color{blue}0.00} & {\color{blue}0.00} & {\color{blue}0.00} & {\color{blue}0.00} & {\color{blue}0.00} & {\color{blue}0.00} & {\color{blue}0.00} & {\color{blue}0.00} & {\color{blue}0.00} & {\color{blue}0.00}   \\
     Demand & & Naive & {\color{red}0.00} & {\color{red}0.00} & {\color{red}0.00} & {\color{red}0.00} & {\color{red}0.00} & {\color{red}0.00} & {\color{red}0.00} & {\color{red}0.00} & {\color{red}0.00} & {\color{red}0.00} & {\color{red}0.00}  \\
     \cline{2-14}
      & Max & EM & 7.25 & {\color{blue}0.78} & {\color{blue}0.36} & {\color{blue}0.47} & {\color{blue}0.23} & {\color{blue}0.28}  & {\color{blue}0.20} & {\color{blue}0.03} & {\color{blue}0.15} & {\color{blue}0.00} & {\color{blue}0.00} \\
      & & Naive & {\color{red}0.00} & 4.00 & 3.33 & 1.90 & 1.57 & 0.81 & 0.91 & 0.61 & 0.63 & 0.21 &  {\color{red}0.00} \\
    \hline
    \end{tabular} } \\
\end{center}
\end{table}

%

\subsection{Sensitivity Analysis} \label{subsec:sens_analysis}

\begin{figure}[h]
\begin{center}
\caption{Visualization of simulated data trip rates and estimated availability.\\}  \label{fig:sensitivity_analysis_sim_data_vis}
\includegraphics[scale=0.25]{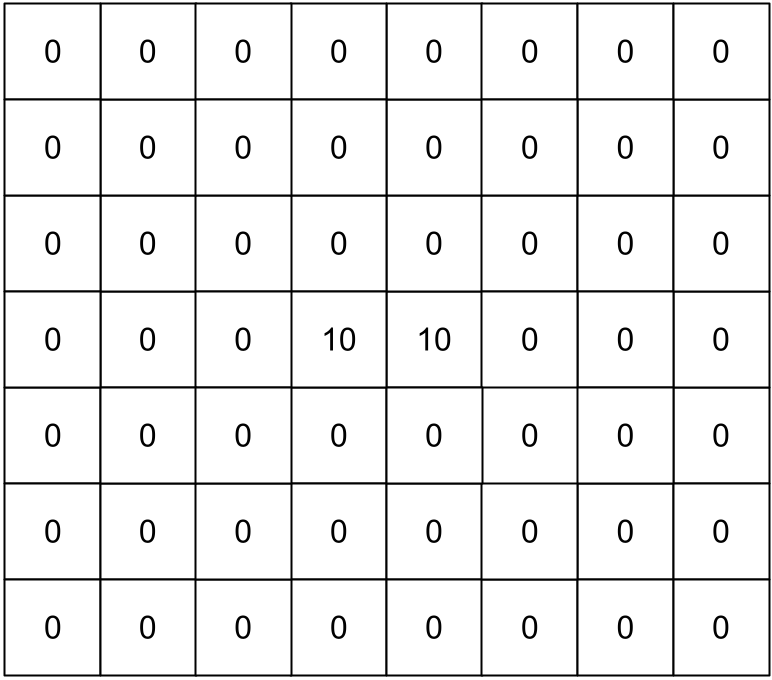}
\includegraphics[scale=0.25]{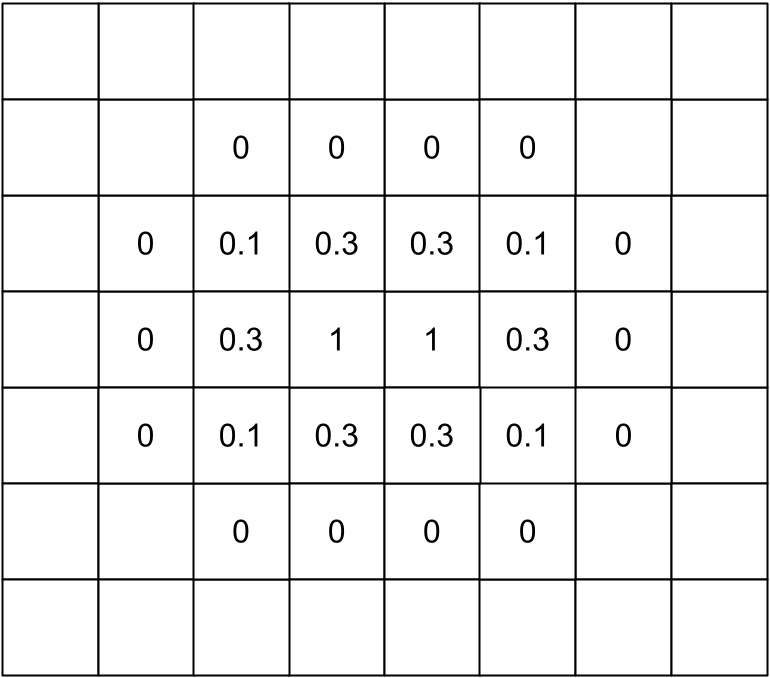}
\end{center}
\end{figure}

When $p=0$ in the computational experiments above, if we initialize $\mu$ based on the observed trip rates rather than uniformly, the EM algorithm performs the same as the Naive algorithm. In this section, we dive deeper into this setting and the sensitivity to the initial values of $\mu$. The EM algorithm in Section~\ref{sec:est} is guaranteed to converge to a local optima, meaning the initial values of the rates $\hat{\mu}$ can impact the returned estimates. We explore the sensitivity of the outcome on the initial starting values on the Providence data and a simple simulated case. For our simulated data, we generate 50 days of trip data with 10 instantaneous trips per hour starting and ending at the same place in two locations 600m apart longitudinally. Our model creates a small grid of 7$\times$8 grid cells around these two active central grid cells, and an initial trip at time zero in both of these cells creates a constant availability of one bike in each of these grid cells for all 50 days. A visualization of the resulting trip rates and estimated availability is given in Figure~\ref{fig:sensitivity_analysis_sim_data_vis}.

For both the Providence data and our simple simulated data, we consider two extremes: (1) setting the rates uniformly so all grid cells start with the same guess of demand $\hat{\mu}_{\text{unif}}$ and (2) setting the rates to the average rate of observed trips in that cell and time period $\hat{\mu}_{\text{trip}}$. While the former corresponds to an uninformed prior on location, the latter ignores any potential travel of users.  To analyze the sensitivity of our algorithm to the initial conditions we consider setting the initial $\hat{\mu}$ values to $\hat{\mu}_{\gamma} = \gamma \hat{\mu}_{\text{unif}} + (1-\gamma) \hat{\mu}_{\text{trip}}$ for $\gamma$ between 0 and 1.

\begin{figure}[h]
\begin{center}
\caption{Visualization of simulated data estimated demand for $\mathbf{\gamma=0}$, $\mathbf{\gamma=0.5}$, and $\mathbf{\gamma=1}$.\\} \label{fig:sensitivity_analysis_gamma_var}
\includegraphics[scale=0.25]{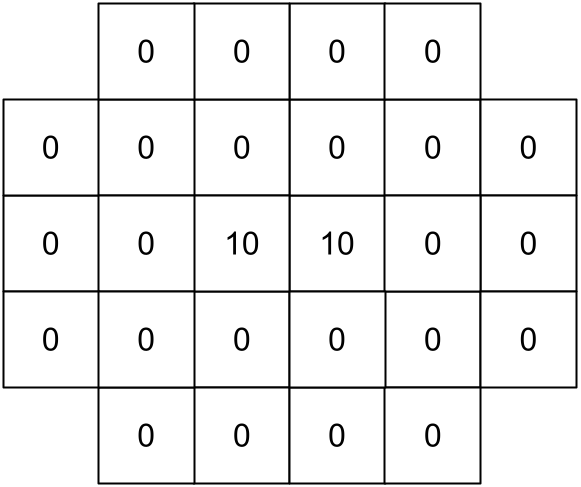}
\includegraphics[scale=0.25]{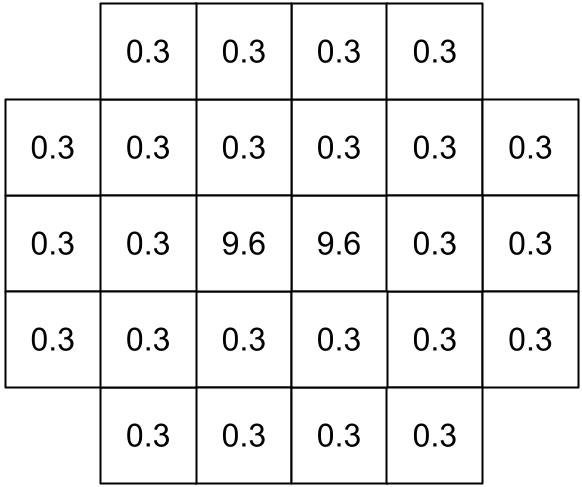}
\includegraphics[scale=0.25]{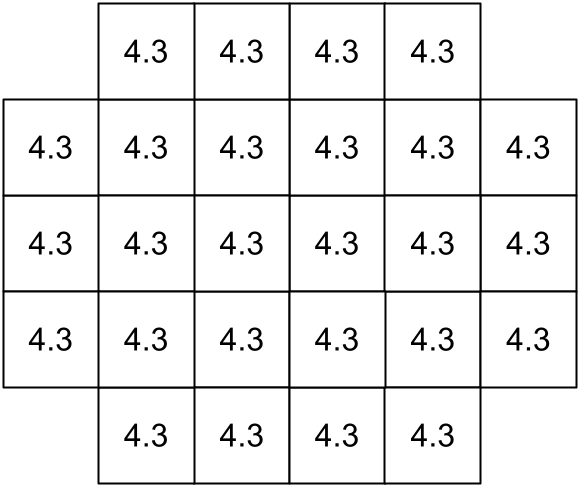}
\end{center}
\end{figure}

\begin{figure}[h]
\centering
\caption{Plot of difference in the estimated demand rates from the trip initialization ($\mathbf{\gamma=0}$) for simple generated data (left) and three months of Providence data (right).} \label{fig:sens_analysis}
\includegraphics[scale=0.5]{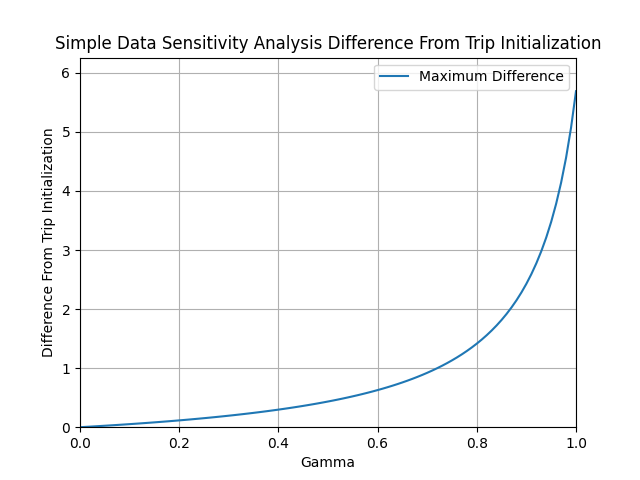}  
\includegraphics[scale=0.5]{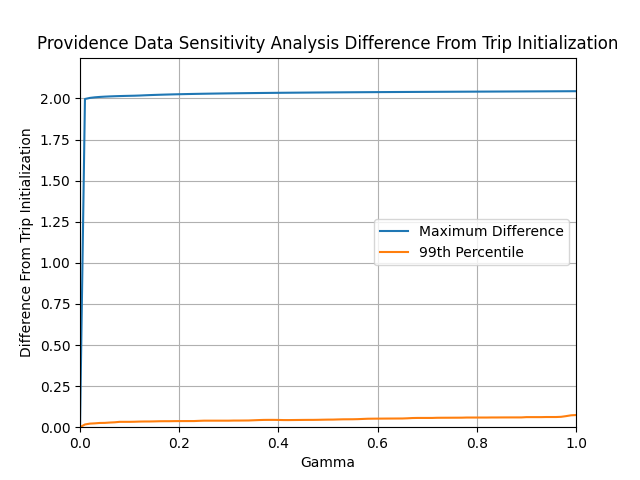}  
\end{figure}

In Figure~\ref{fig:sens_analysis}, we plot the largest difference between the $\mu_{\text{trip}}$ initialization and the $\mu_{\gamma}$ initialization versus $\gamma$. In the simulated data, we observe that the largest difference increases steadily as $\gamma$ increases, shown in Figure~\ref{fig:sensitivity_analysis_gamma_var}. This is expected because we have a limited distribution on availability. As $\gamma$ increases, the estimated demand spreads from the central cells to the surrounding cells until uniform at $\gamma=1$. Hence, the estimated demand is highly sensitive to the initialization.

By contrast, in the Providence data, we observe an immediate increase to a largest difference of around $2$, and a subsequent stabilization of the largest difference at this value. The 99th percentile stays relatively low throughout, and the median difference was observed to be 0 for all $\gamma$. To generate these results, we did not include grid cells in which estimated $\hat{\alpha}$ is less than $10^{-2}$, the same threshold used in our application. We consider these cells not to have enough availability to infer demand. Note that for regions that did not meet this threshold, the maximum observed difference between $\hat{\mu}_{\text{trip}}$ and $\hat{\mu}_{\text{unif}}$ was $103$. This is caused by numerical instability in the division by small $\pi$ and $\hat{\alpha}$ values in the EM algorithm iterations.

Overall, these results indicate that when we have very limited availability, the results are more sensitive to the initialization since we are unable to discern where users may be arriving from. However, with more a variable and realistic distribution of availability the results are far more robust to our initialization. Given the results in Section~\ref{subsec:comp_results}, in which the algorithm does better at predicting bordering demand but worse at estimating demand in the cluster centers, we choose to initialize with $\hat{\mu}_{\text{unif}}$ in our implementation to err on the side of distributed demand.

\section{Application and Providence Results} \label{sec:ex}

\begin{figure}[h]
\begin{center}
\caption{Labeled screenshot of the application visualizing Providence demand data.} \label{fig:app_screenshot}
\includegraphics[scale=0.28]{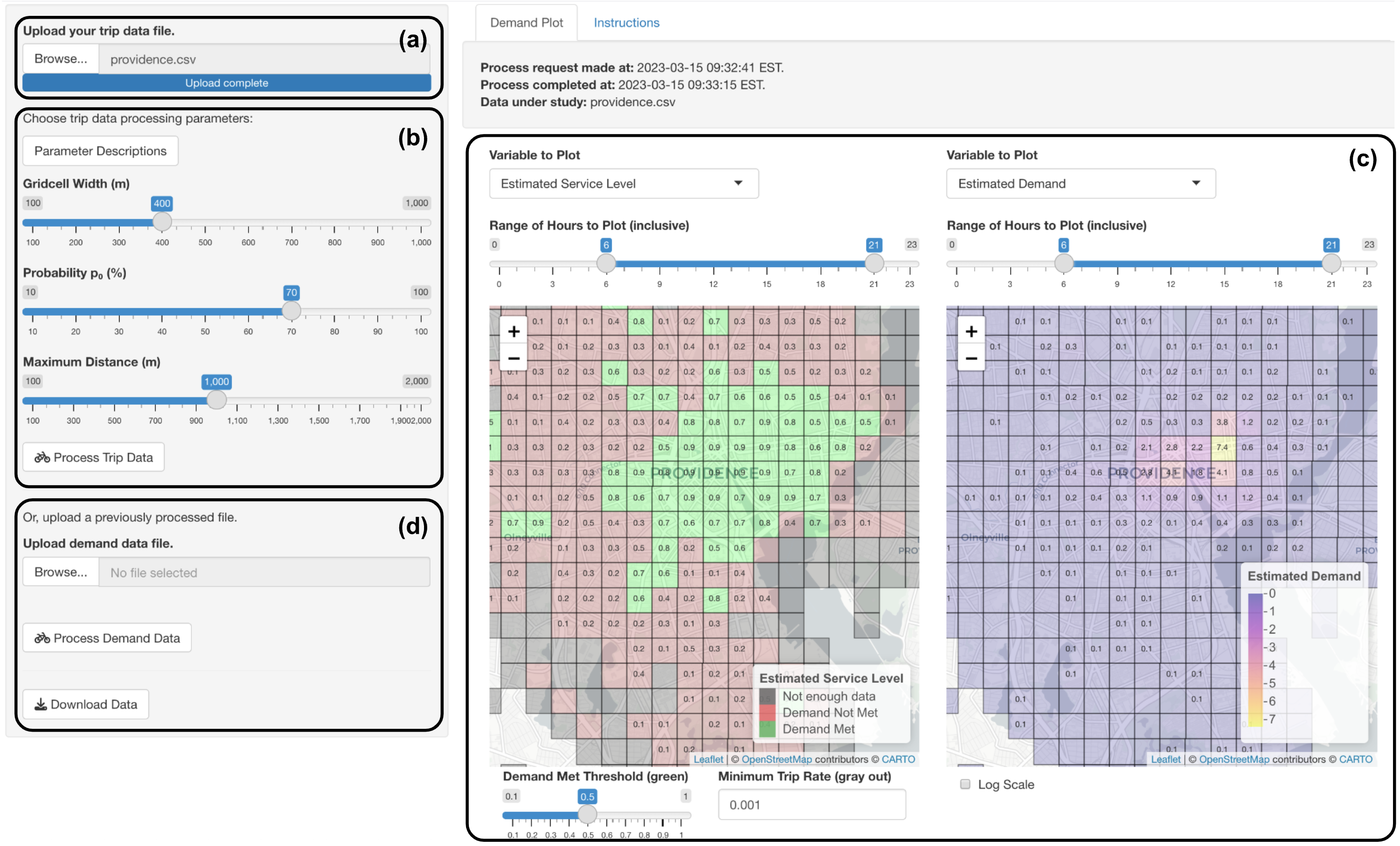}  
\end{center}
\end{figure}

To create a user-friendly interaction with our model and to visualize the results, we host a live R Shiny application that communicates with our python demand model through the reticulate package. We display its labeled interface in Figure~\ref{fig:app_screenshot}. The app allows users to easily upload their trip data in (a), select the demand model parameter values of choice in (b) (grid cell width, probability $p_0$, maximum user distance $\mathrm{dist}_{max}$), and run the model for demand visualization with a click of a button. Once processed, we display side-by-side colored leaflet maps in (c) that allow users to analyze and compare the findings of our model in their city. From a drop-down menu above them, users can choose to visualize the estimated demand rates, the estimated or observed availability, the rate at which trips took place, and a summary visualization that categorizes service levels to highlight where the rate of observed trips is much lower than the estimated demand. These findings allow city planners to determine regions of unmet demand due to low availability not directly inferable from the raw trip data. Users can then download the estimated model data for further data analysis or re-upload downloaded data for re-visualization in (d). Our application can be accessed at \href{https://alicejpaul.shinyapps.io/shared-mobility/}{https://alicejpaul.shinyapps.io/shared-mobility/}. A local version of the app can also be downloaded through the github repository (\href{https://github.com/KyFlynn/shared-mobility-research}{https://github.com/KyFlynn/shared-mobility-research}).

For Providence data with approximately 100k trips (200k data points) spanning three months distributed over a 276km$^2$ ($\sim$107mi$^2$), with default parameter settings (grid cell width of 400m, probability $p_0$ of 0.7, maximum walking distance of 1km) our application takes approximately 30s locally and 2min on our live app.\footnote{Locally: Apple 2022 16" Macbook Pro, M1 Pro chip: 32GP unified memory; 10-core CPU (8 performance, 2 efficiency); 16-core GPU; 16-core Neural Engine; 200Gbps memory bandwidth.} In Table~\ref{tab:runtime_table}, we report the runtime of our app both locally and on the published Shiny app for different grid cell sizes and numbers of trips, keeping all other input parameters equal. Entries without data in the website column mean the website ran out of memory running the demand model. For large amounts of data or for high granularity of estimated demand, we recommend downloading the application from the \href{https://github.com/KyFlynn/shared-mobility-research}{github repo} and running it locally.

\begin{table}[h]
\centering
\caption{\label{tab:runtime_table}Runtime of our demand model locally versus on our public application for different grid cell sizes and number of trips in input trip data.} 
\begin{tabular}{|c|c|c|c|}
\hline
\textbf{Grid Cell Width (m)} & \textbf{Number of Trips} &\textbf{Local Runtime (s)} & \textbf{Website Runtime (s)}\\
\hline
 & 100k & 107 & -\\
\cline{2-4}
200 & 200k & 460 & -\\
\cline{2-4}
& 300k & 871 & -\\
\cline{1-4}
 & 100k & 34 & 144\\
\cline{2-4}
400 & 200k & 63 & 598\\
\cline{2-4}
 & 300k & 87 & -\\
\cline{1-4}
 & 100k & 16 & 44\\
\cline{2-4}
600 & 200k & 32 & 130\\
\cline{2-4}
 & 300k & 46 & 293\\
\hline
\end{tabular}
\end{table}

\subsection{Example Results}

\begin{figure}
\centering
\caption{Visualization of Providence data summer 2019 results in the application.}
\label{fig:pvd_results}
    \begin{subfigure}[b]{0.45\textwidth}
         \centering
         \includegraphics[width=\textwidth]{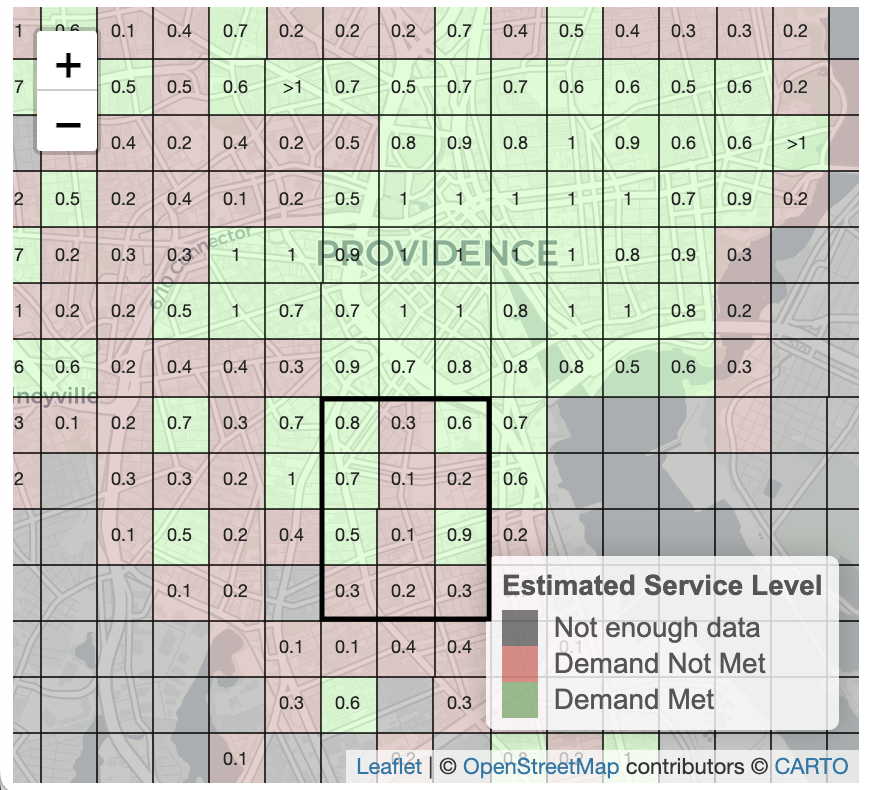}
         \caption{Estimated Service Level.}
     \end{subfigure}
     \hfill
    \begin{subfigure}[b]{0.45\textwidth}
         \centering
         \includegraphics[width=\textwidth]{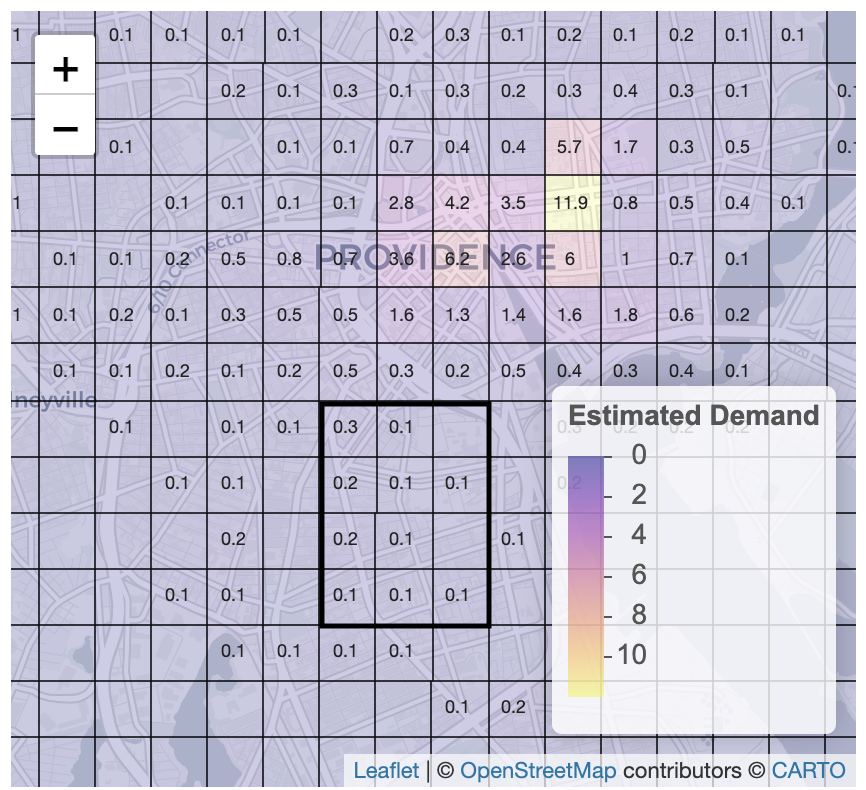}
         \caption{Estimated Demand Rates.}
     \end{subfigure}
     \hfill
    \begin{subfigure}[b]{0.45\textwidth}
         \centering
         \includegraphics[width=\textwidth]{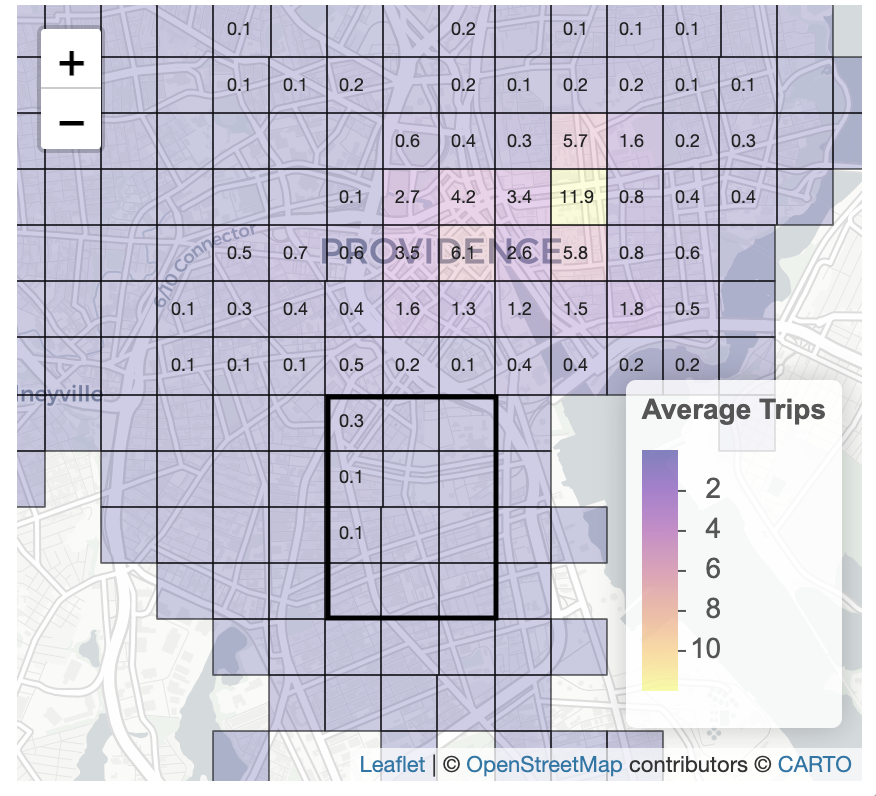}
         \caption{Average Number of Trips.}
     \end{subfigure}
     \hfill
     \begin{subfigure}[b]{0.45\textwidth}
         \centering
         \includegraphics[width=\textwidth]{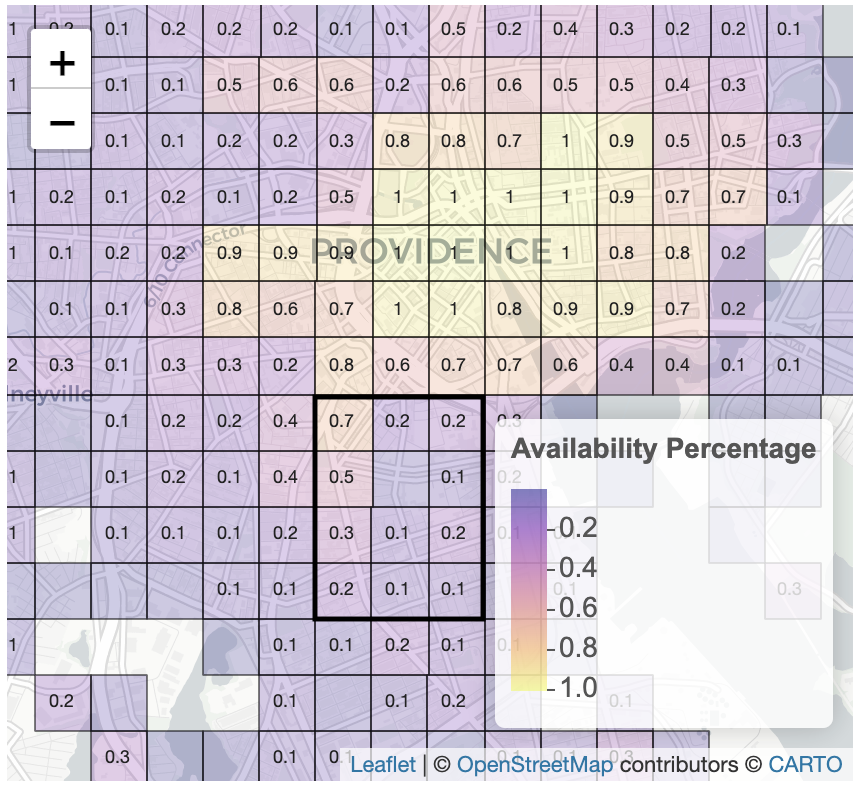}
         \caption{Percentage of Time Available.}
     \end{subfigure}
\end{figure}

We now present results from scooter data from Providence, RI for June 1, 2019 to August 31, 2019. In Figure~\ref{fig:pvd_results}, the plots display an estimation of the service level, the estimated demand rates, the percentage of time at least one scooter is available, and the average number of trips observed for noon to 3pm. 
In Figure~\ref{fig:pvd_results}~(a), we highlight cells where the estimated demand is at least twice the observed trip rate in red. The plots show that the observed number of trips reflects the estimated demand in areas with high availability but that there are also many cells with low availability and some estimated demand. For example, in the highlighted region on the four plots, we observe that over half of these cells have low service levels (indicated by red squares) with very low availability and few observed trips. This region is part of the overall `South' region in which operators regularly fail to meet the 5\% of trips requirement. However, the estimated demand shows that when there is availability, we observe some user trips. While the estimated demand rates are much lower than in the high frequency areas highlighted in yellow in Figure~\ref{fig:pvd_results} (b), it could be the case that more consistent availability will increase usage rates. 

Overall, the plots also indicate why the operators may be having trouble meeting the service requirements by highlighting areas with low demand and/or low availability. To increase equity in access and the overall service level, we can use the plots to identify areas with the highest estimated demand with low service levels as areas to target with service expansion. Additionally, there are some areas without enough data to estimate demand. This indicates we need further exploration to better understand demand. 



Our github repository has a test data set available for people to try out the application. We downloaded the trip data from Kansas City, MO's Microtransit Network \citep{kansasMicro}. We first filter the data to only consider trips from May 1, 2021 to May 6, 2021. The public data does not contain any operator rebalancing. To account for this, we assume perfect rebalancing each day. At the end of each day, bikes are removed and the minimum number needed to ensure the observed trips the next day are feasible are relocated. Our R script for processing the data is also available in our repository.

\section{Future Work} \label{sec:future}


This paper presents a flexible and interpretable framework to estimate spatial-temporal demand from censored data that bridges the gap between the docked and dockless shared mobility literature. Some possible extensions to this model include incorporating the direction of travel and the built environment into our user choice model. Additionally, we can extend the model to allow the rates to depend on other factors such as the day of the week, season, and weather. Lastly, the results provide insight into user behavior and can be used to inform future planning decisions such as redistribution and identifying areas with unmet demand. For example, planners could use the estimated stochastic process to simulate user behavior and test out different settings such as different distributions for availability or a different fleet sizes. This focus on prescriptive analytics using the estimated model is a promising direction.

\section{Acknowledgements}
The authors would like to acknowledge principal planner Alex Ellis and curbside administrator Liza Farr with the City of Providence for their help defining our model and framework and their insight into the data and results.

\bibliographystyle{plainnat} 
\bibliography{em_estimation.bib}

\end{document}